\documentclass[11pt]{cernrep}
\usepackage{graphicx}
\usepackage{cite,./mcite}
\usepackage{epsfig}
\usepackage{amssymb}
\long\def\symbolfootnote[#1]#2{\begingroup%
\def\thefootnote{\fnsymbol{footnote}}\footnote[#1]{#2}\endgroup}

\begin{document}
\title{Diffraction at TOTEM}
\author{G.~Antchev$^{1}$,
P.~Aspell$^{1}$, V.~Avati$^{1,9}$, M.G.~Bagliesi$^{5}$,
V.~Berardi$^{4}$, M.~Berretti$^{5}$, U.~Bottigli$^{5}$,
M.~Bozzo$^{2}$,
E.~Br\"{u}cken$^{6}$, A. Buzzo$^{2}$, F.~Cafagna$^{4}$, M.~Calicchio$^{4}$,
M.G.~Catanesi$^{4}$,
P.L.~Catastini$^{5}$, R.~Cecchi$^{5}$,
M.A.~Ciocci$^{5}$,
M.~Deile$^{1}$, E.~Dimovasili$^{1,9}$,
K.~Eggert$^{9}$, V.~Eremin$^{\ast\ast}$, F.~Ferro$^{2}$,
F.~Garcia$^{6}$, S.~Giani$^{1}$, V.~Greco$^{5}$, J.~Heino$^{6}$, T.~Hild\'en$^{6}$,
J.~Ka\v{s}par$^{1,7}$, J.~ Kopal$^{1,7}$, V.~Kundr\'{a}t$^{7}$, K.~Kurvinen$^{6}$, S.~Lami$^{5}$, G.~Latino$^{5}$,
R.~Lauhakangas$^{6}$, E.~Lippmaa$^{8}$,
M.~Lokaj\'{\i}\v{c}ek$^{7}$, M.~Lo~Vetere$^{2}$, F.~Lucas~Rodriguez$^{1}$, M.~Macr\'{\i}$^{2}$, G.~Magazz\`{u}$^{5}$,
M.~Meucci$^{5}$, S.~Minutoli$^{2}$,
H.~Niewiadomski$^{1,9}$, E.~Noschis$^{1}$,
G.~Notarnicola$^{4}$,
E.~Oliveri$^{5}$, F.~Oljemark$^{6}$, R.~Orava$^{6}$, M.~Oriunno$^{1}$, K.~\"{O}sterberg$^{6,}$\symbolfootnote[3]{\hspace{1mm} corresponding author: Kenneth \"Osterberg (kenneth.osterberg@helsinki.fi)}$\,$,
P.~Palazzi$^{1}$,  E.~Pedreschi$^{5}$,
J.~Pet\"{a}j\"{a}j\"{a}rvi$^{6}$, M.~Quinto$^{4}$,
E.~Radermacher$^{1}$, E.~Radicioni$^{4}$,
F.~Ravotti$^{1}$, G.~Rella$^{4}$, E.~Robutti$^{2}$,
L.~Ropelewski$^{1}$, G.~Ruggiero$^{1}$, A.~Rummel$^{8}$,
H.~Saarikko$^{6}$, G.~Sanguinetti$^{5}$, A.~Santroni$^{2}$,
A.~Scribano$^{5}$, G.~Sette$^{2}$, W.~Snoeys$^{1}$, F.~Spinella$^{5}$,
P.~Squillacioti$^{5}$, A.~Ster$^{\ast}$, C.~Taylor$^{3}$, A.~Trummal$^{8}$,
N.~Turini$^{5}$, J.~Whitmore$^{9}$, J.~Wu$^{1}$}
\institute{$^{1}$CERN, Gen\`{e}ve, Switzerland,\\
$^{2}$Universit\`{a} di Genova and Sezione INFN, Genova, Italy,\\
$^{3}$Case Western Reserve University, Dept. of Physics, Cleveland, OH, USA,\\
$^{4}$INFN Sezione di Bari and Politecnico di Bari, Bari, Italy,\\
$^{5}$INFN Sezione di Pisa and Universit\`{a} di Siena, Italy,\\
$^{6}$Helsinki Institute of Physics and Department of Physics,
University of Helsinki, Finland,\\
$^{7}$Institute of Physics of the Academy of Sciences of the Czech Republic,
Praha, Czech Republic,\\
$^{8}$National Institute of Chemical Physics and Biophysics NICPB, Tallinn,
Estonia.\\
$^{9}$Penn State University, Dept. of Physics, University Park, PA, USA.\\
$^{\ast}$Individual participant from MTA KFKI RMKI, Budapest, Hungary.\\
$^{\ast\ast}$On leave from Ioffe Physico-Technical Institute,
Polytechnicheskaya Str. 26, 194021 St-Petersburg, Russian Federation.}
\maketitle
\begin{abstract}
The TOTEM experiment at the LHC measures the total proton-proton cross section with the luminosity-independent method and the elastic
proton-proton
cross-section over a wide $|t|$-range. It also performs a comprehensive study of diffraction, spanning from cross-section measurements of individual diffractive processes to the analysis of their
event topologies. Hard diffraction will be studied in collaboration with CMS taking advantage of the large common rapidity coverage for charged and neutral particle detection and the large variety of trigger possibilities even at large luminosities. TOTEM will take data under all LHC beam conditions including standard high luminosity runs to maximize its physics reach. This contribution describes the main features of the TOTEM physics programme including measurements to be made in the early LHC runs. In addition, a novel scheme to extend the diffractive proton acceptance for high luminosity runs by installing proton detectors at IP3 is described.
\end{abstract}

\section{Introduction}
\label{sec:intro}
\begin{figure}[h!]
\begin{center}
\epsfig{file=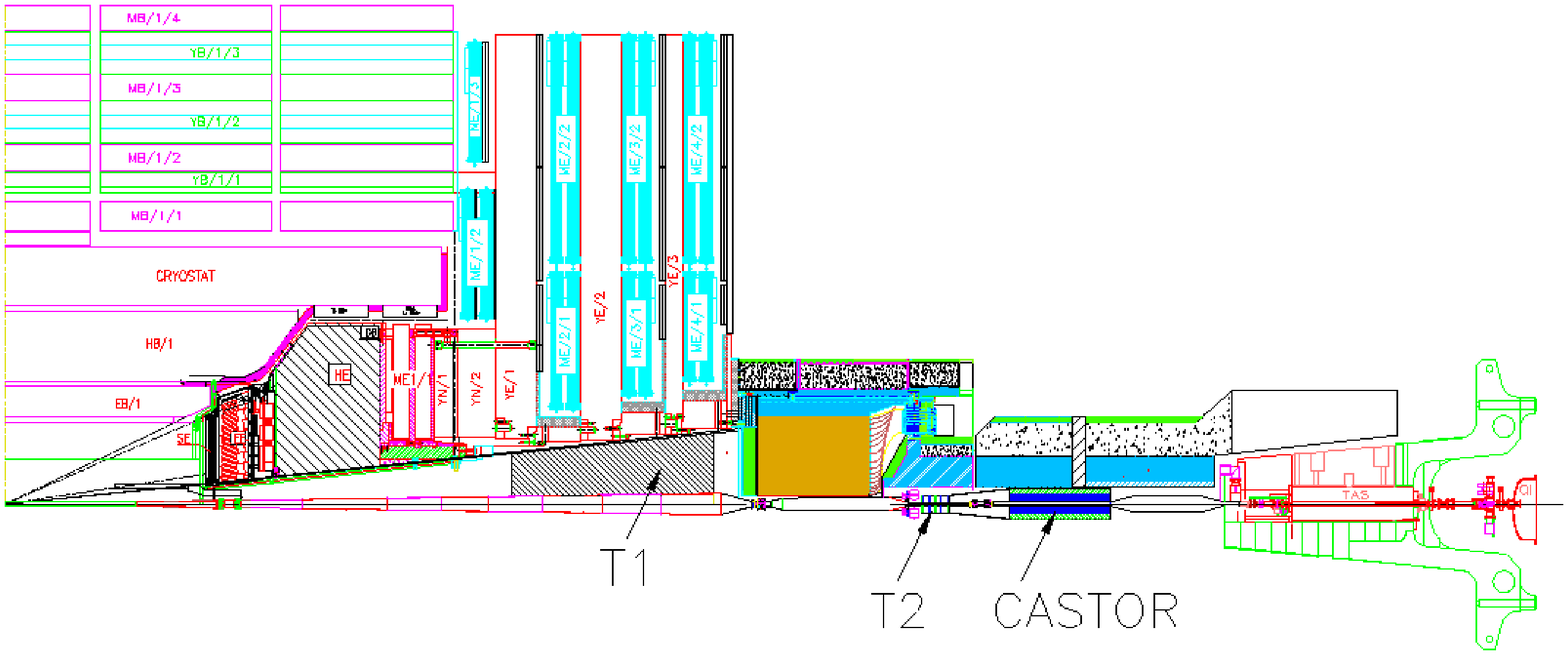,width=13.75cm}
\epsfig{file=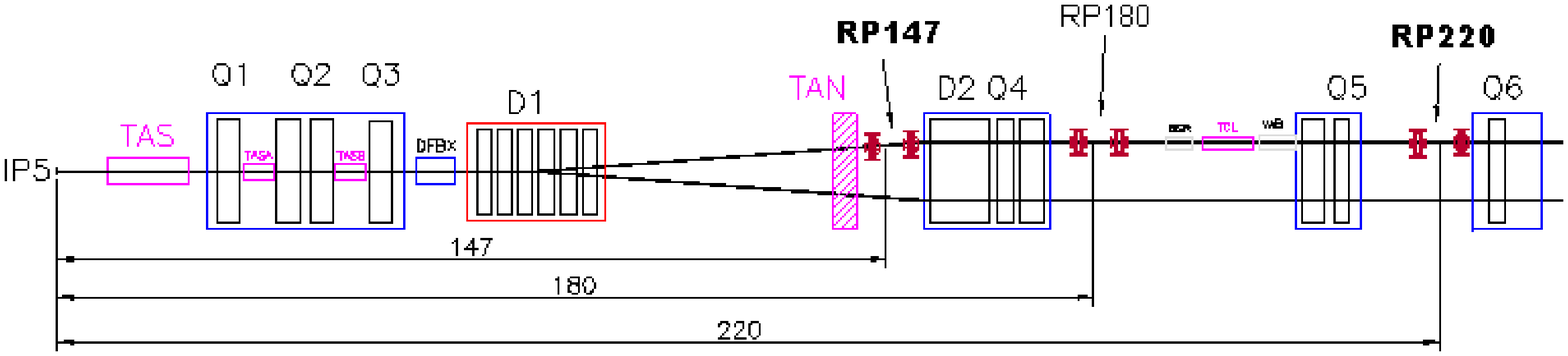,width=13.75cm}
\caption{Top: TOTEM forward telescopes T1 and T2 embedded in the CMS experiment
together with the CMS forward calorimeter CASTOR. Bottom: LHC beam line on one side of interaction point IP5 and TOTEM Roman Pot stations at distances of about 147\,m (RP147) and 220\,m (RP220).
RP180 at 180\,m is another possible location but presently not equipped.}
\label{fig_apparatus}
\end{center}
\end{figure}

The TOTEM experiment~\cite{TOTEMTDR} is dedicated to the
total proton-proton ($pp$) cross-section measurement using the
luminosity-independent method, which requires
a detailed measurement of the elastic scattering rate down to a squared four-momentum transfer of
$-t \sim p^2\Theta^2 \sim \rm 10^{-3}\,GeV^{2}$ together with the measurements of the total inelastic
and elastic rates.
Furthermore, by
studying elastic scattering with momentum transfers up to 10 GeV$^2$,
and via a comprehensive study of diffractive processes -- partly in cooperation
with CMS~\cite{physicstdr}, located at the same interaction point, TOTEM's physics programme aims at a deeper understanding of
the proton structure.
To perform these measurements, TOTEM requires a good acceptance for particles
produced at small and even tiny angles with respect to the beams. TOTEM's
coverage in the pseudo-rapidity range of $3.1 \le |\eta| \le 6.5$
($\eta = -\ln \tan \frac{\theta}{2}$) on both sides of
the interaction point (IP) is accomplished by two telescopes, T1 and T2 (Figure~\ref{fig_apparatus}, top), for the detection of charged particles with emission angles between a few and about hundred milliradians. This is complemented by detectors in special movable beam-pipe
insertions -- so called Roman Pots (RP) -- placed at about
147\,m and 220\,m from the IP, designed to detect elastically or diffractively scattered protons at merely a few millimeter from
the beam center corresponding to emission angles down to a few microradians (Figure~\ref{fig_apparatus}, bottom).

For the luminosity-independent total cross-section measurement, TOTEM has to
reach the lowest possible $|t|$ values in elastic $pp$ scattering. Elastically scattered protons close to the beam can be detected downstream
on either side of the IP
if the displacement at the detector location is large enough and if the
beam divergence at the IP is
small compared to the scattering angle.
To achieve these conditions special LHC optics with high beta value at the IP ($\beta^*$) are required:
the larger the $\beta^*$, the smaller the beam divergence ($\sim 1/\sqrt{\beta^*}$) will be. Two optics are proposed: an ultimate one with $\beta^* = 1540$\,m
and another one, possibly foreseen for 2009, with $\beta^* = 90$\,m. The latter uses the standard injection
optics ($\beta^*$ = 11 m) and beam conditions typical for early LHC running: zero degree
crossing-angle and consequently at most 156 bunches together with a low
number of protons per bunch.

The versatile physics programme of TOTEM requires different running scenarios that
have to be adapted to the LHC commissioning and operation
in the first years. A flexible
trigger can be provided by the two telescopes and the Roman Pot detectors.
TOTEM will take data under all optics conditions,
adjusting the trigger schemes to
the luminosity. The DAQ will allow trigger rates up to a few kHz without
involving a higher level trigger. The high-$\beta^*$ runs (Table~\ref{tab_par}) with 156 bunches, zero degree
crossing-angle and maximum luminosity between $10^{29}$ and
$10^{30}\,\rm cm^{-2}s^{-1}$, will concentrate on low-$|t|$ elastic scattering,
total cross-section, minimum bias physics and soft diffraction.
A large fraction of forward protons will be detected even at the lowest $\xi$
values.
Low-$\beta^*$ runs (Table~\ref{tab_par}) with more bunches and higher
luminosity ($10^{32}$ -- $10^{34}\,\rm cm^{-2}s^{-1}$)
will be used for large-$|t|$ elastic scattering and diffractive studies with
$\xi>0.02$. Hard diffractive events come within reach. In addition, early low $\beta^*$ runs will provide first opportunities for measurements of soft  diffraction at LHC energies and for studies of forward charged multiplicity.

\begin{table}[ht]
\begin{center}
\begin{tabular}{|cccccc|}\hline
$\beta^*$ [m] & $k$ & $N$/10$^{11}$ & $\mathcal{L}$ [cm$^{-2}$s$^{-1}$] &
$|t|$-range [GeV$^2$] @ $\xi=0$ & $\xi$-range \\ \hline
1540 & 43 $\div$ 156 & 0.6 $\div$ 1.15 & 10$^{28}$ $\div$ 2 $\cdot$ 10$^{29}$ &
0.002 $\div$ 1.5  & $<$ 0.2 \\
90   & 156 & 0.1 $\div$ 1.15 & 2 $\cdot$ 10$^{28}$ $\div$ 3 $\cdot$ 10$^{30}$ &
0.03 $\div$ 10  & $<$ 0.2\\
11 & 43 $\div$ 2808 & 0.1 $\div$ 1.15 & $\sim$ 10$^{30}$ $\div$ 5 $\cdot$ 10$^{32}$ &
0.6 $\div$ 8  & 0.02 $\div$ 0.2 \\
0.5 $\div$ 3   & 43 $\div$ 2808 & 0.1 $\div$ 1.15 & $\sim$ 10$^{30}$ $\div$ 10$^{34}$ &
2 $\div$ 10  & 0.02 $\div$ 0.2\\
\hline
\end{tabular}
\caption{Running scenarios at different LHC optics ($k$: number of bunches, $N$: number of particles per bunch, $\mathcal{L}$: estimated luminosity). The $|t|$ ranges for elastically scattered protons correspond to the
$\ge 50$\% combined RP147 and RP220 acceptance.}
\label{tab_par}
\end{center} \vspace*{-0.8cm}
\end{table}

In the following, after a brief description of the TOTEM detectors and the principles of proton detection, the main features of the TOTEM physics programme will be given. This will be followed by a description of the early physics programme. Finally the novel idea of proton detection at IP3 will be presented. A detailed technical description of the TOTEM experiment can be found in Ref.~\cite{LHCarticle}.
\section{TOTEM detectors and performance}
\label{sec:det}
\subsection{Inelastic detectors}
The measurement of the inelastic rate requires identification of all beam-beam events with detectors capable to trigger and reconstruct the interaction vertex. The main requirements of these detectors are:
\begin{itemize}
 \item to provide a fully inclusive trigger for minimum bias and diffractive events, with minimal
losses at a level of a few percent of the inelastic rate;
 \item to enable the reconstruction of the primary vertex of an event, in order to disentangle beam-beam
events from the background via a partial event reconstruction.
\end{itemize}
These requirements are fulfilled by the T1 telescope (centered at $z$ = 9\,m), consisting of Cathode Strip Chambers (CSC) and T2 telescope (centered at $z$ = 13.5\,m) exploiting Gas Electron Multipliers (GEM). The $\eta$ coverage of T1 and T2 is $3.1 \le |\eta| \le 4.7$ and $5.3 \le |\eta| \le 6.5$, respectively. Each T1 telescope arm consists of five planes made up of six trapezoidal formed CSC's with a spatial resolution of $\sim$ 1 mm. Each T2 telescope arm consists of 20 semicircular shaped triple-GEM detectors with a spatial resolution of $\sim$ 100 $\mu$m in the radial direction and a inner radius that matches the beam-pipe. Ten aligned detectors mounted back-to-back are combined to form one T2 half arm on each side of the beam-pipe. For charged particles with momenta typical of particles produced within the detector acceptances in inelastic events, the particle $\eta$ can be determined with a precision that increases with $|\eta|$ and is between 0.02 and 0.06 in T1 and between 0.04 and 0.1 in T2. The corresponding azimuthal angle resolution for both detectors is $\sim$ 1$^o$. The magnetic field at the detector locations is too weak to allow for a momentum determination for the charged particles.
The primary vertex can be reconstructed with a precision of $\sim$ 1.5 cm in the radial direction and $\sim$ 20 cm in the beam direction in presence of the CMS magnetic field.
Vertex resolutions one order of magnitude better can be achieved running with the CMS magnetic field
switched off.

\subsection{Proton detectors}
To measure elastically and diffractively scattered protons with high acceptance requires the reconstruction of the protons tracks by ``trigger capable" detectors moved as close as $\sim$ 1 mm from the center of the outgoing beam. This is obtained with
two RP stations installed, symmetrically on both sides of IP5, at a distance of $\sim$ 147\,m and $\sim$ 220\,m from IP5. These positions are given by an interplay between the development of the special TOTEM
optics and the constraints given by the LHC accelerator elements.
Each RP station is composed of two units at a distance of several meters.
This large lever arm allows local track reconstruction and a fast trigger selection based on the track angle.
Each unit consists of three pots, two approaching the beam vertically from the top and the bottom and one
horizontally to complete the acceptance for diffractively scattered protons, in particular for the low $\beta^*$ optics. Furthermore,
the overlap of the detector acceptance in the horizontal and vertical pots is vital for the relative alignment of the three pots via common particle tracks. The position of the pots with respect to the beam is given by Beam Position Monitors mechanically fixed to all three pots in one unit. Each pot contains a stack of 10 planes of silicon strip "edgeless" detectors with half with their strips oriented at an angle of $+45^o$ and half at an angle of $-45^o$ with respect to the
edge facing the beam. These detectors, designed
by TOTEM with the objective of reducing the insensitive area at the
edge facing the beam to only a few tens of microns, have a spatial resolution of $\sim$ 20 $\mu$m. High efficiency up to the
physical detector border is essential in view of maximizing the elastic and
diffractive proton acceptances. For the same reason, the pots' stainless steel bottom foil that faces the beam has been reduced to a thickness of 150 $\mu$m.

\subsection{Proton detection}
The transverse displacement $(x(s), y(s))$ of an elastically or diffractively
scattered proton
at a distance $s$ from the IP is related to its origin $(x^*, y^*, 0)$,
scattering angles $\Theta^*_{x,y}$ and fractional momentum loss $\xi$ (= $\Delta p/p$) value at the IP via the optical
functions $L$ and $v$, and the dispersion $D$:
\begin{eqnarray}
x(s) = v_x(s)\cdot x^* + L_{x}(s) \cdot \Theta^*_x + \xi \cdot D(s) \, \, \,
& {\rm and} & \, \, \, y(s) = v_y(s)\cdot y^* + L_{y}(s) \cdot \Theta^*_y \label{eqn_coord}
\end{eqnarray}
$L,v$ and $D$ determining the
explicit path of the proton through the LHC elements,
depend mainly on the position along the beam line
i.e. on all the elements
traversed before reaching that position and their settings, which is a optics dependent repetition, and hence the RP acceptance for leading
protons will depend on the optics. The allowed minimum
distance of a RP to the beam center on
one hand being proportional to the beam size ($(10-15) \cdot \sigma_{x(y)}(s)$) as well as constraints imposed by the beam-pipe or beam screen
size on the other hand will determine the proton acceptance of a
RP station.

\begin{figure}[h!]
\begin{center}
\mbox{
\epsfig{file=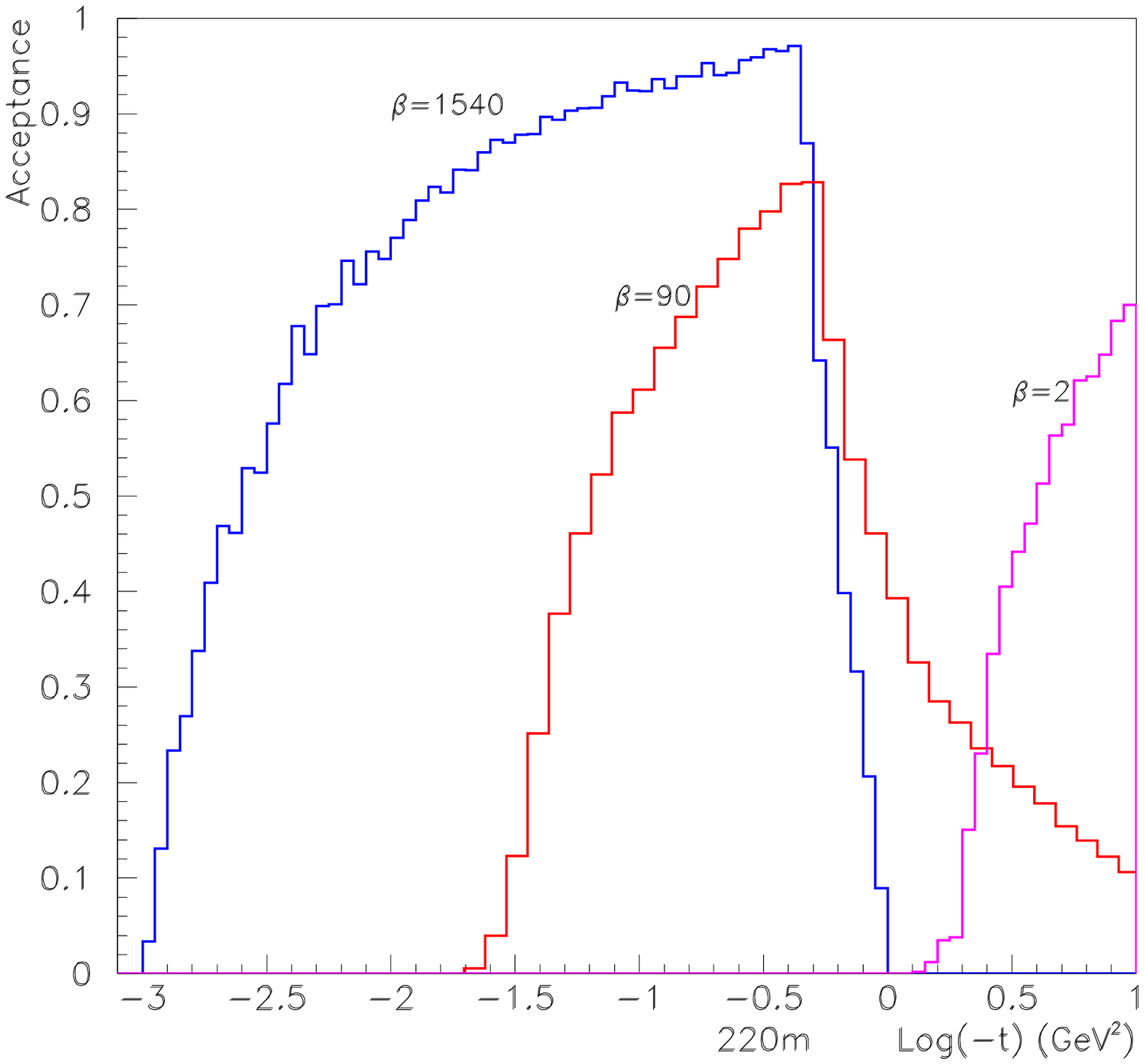,height=6.4cm}
\epsfig{file=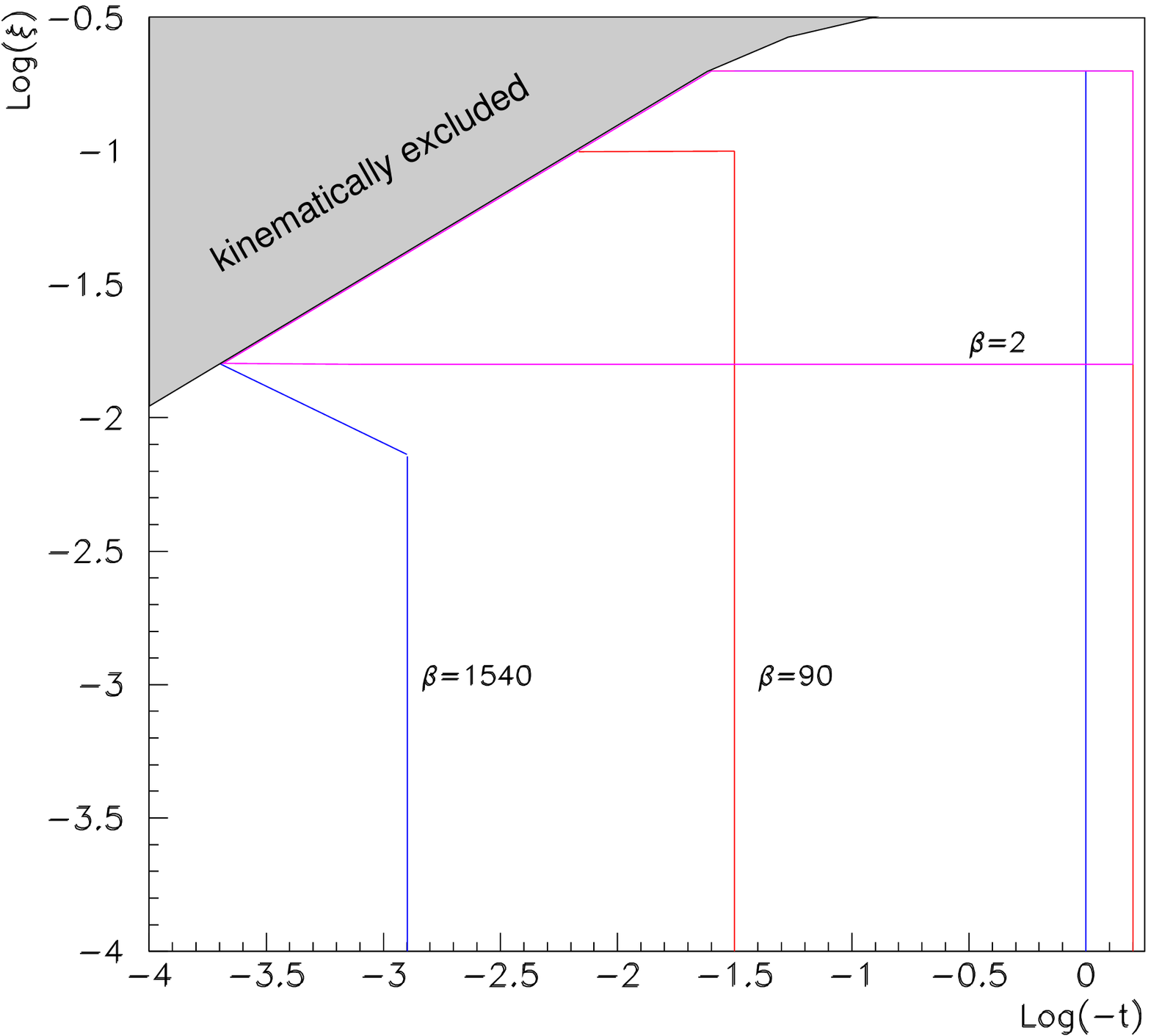,height=6.4cm}
}
\vspace*{-0.5cm}
\end{center}
\caption{Left: RP220 $\log_{10}|t|$ acceptance for elastically scattered
protons at different optics configurations.
Right: contour lines of 10 \% acceptance for RP220 in $\log_{10} |t|$ and $\log_{10} \xi$ for diffractively
scattered protons at different optics configurations.
}
\label{fig_physperf_accept}
\end{figure}

The complementarity of the
acceptances for different optics configurations
is shown in Figure~\ref{fig_physperf_accept}. The TOTEM-specific optics with $\beta^{*} = 1540\,$m (blue graphs in
Figure~\ref{fig_physperf_accept})
is particularly optimized for accepting protons down to very low $|t|$-values. For the diffractive case all kinematically allowed values of $\xi$ are accepted. With the $\beta^{*} = 90\,$m optics (red graphs in
Figure~\ref{fig_physperf_accept}),
diffractive scattered protons are still
accepted independently of their $\xi$-value, but
the $t$-acceptance is reduced compared to $\beta^{*} = 1540\,$m optics.
With the standard high
luminosity optics ($\beta^{*}$ = 0.5 $\div$ 3\,m, magenta graphs in Figure~\ref{fig_physperf_accept}) elastically scattered protons can only be detected
at very large $|t|$ and diffractively scattered protons are accepted independently of their $t$ value in the horizontal pots for $\xi$ values above 2 \%.

\begin{table}[ht]
\begin{center}
\vspace*{-0.30cm}
\begin{tabular}{|c|ccc|}\hline
    & $\beta^*$ = 0.5 -- 3 m & $\beta^*$ = 90 m  &  $\beta^*$ = 1540 m \\ \hline
$\sigma(\xi)$ & 0.001 $\div$ 0.006 & $\sim$ 0.0015 (w CMS vtx) & 0.002 $\div$ 0.006 $^\ddagger$ \\
& & $\sim$ 0.006 (w/o CMS vtx)& \\ \hline
$\sigma(t)$ [GeV$^2$] & (0.3$\div$0.45)~$\sqrt{|t|}$ & $\sigma(t_y)$ $\sim$ 0.04~$\sqrt{|t_y|}$ & $\sim$ 0.005~$\sqrt{|t|}$ \\ \hline
$\sigma(M)$ [GeV] in & (0.02 $\div$ 0.05)~$M$ & $<$ 18 for R $>$ 0.6 (w CMS vtx)  & $\sim$ 20~$M^b$ $^\ddagger$\\
central diffraction & & $<$ 80 for R $>$ 0.6 (w/o CMS vtx)  & $b$ = 0.17 for R = 0.5 $\div$ 1 \\
\hline
\end{tabular}
\caption{Summary of resolutions for the RP220 proton
reconstruction at different optics configurations. ``w CMS vtx'' and ``w/o CMS vtx'' refers to whether vertex position information from CMS is available or not (relevant for $\beta^*$ = 90 m), R = $\xi_{lower}/\xi_{higher}$
to the momentum loss symmetry between the two outgoing protons and ``$\ddagger$'' to reconstruction using also RP147.}
\label{tab_pres}
\end{center} \vspace*{-0.8cm}
\end{table}

The reconstruction of the proton kinematics is optics dependent. The main resolutions are given in Table~\ref{tab_pres}. More details can be found in Ref.~\cite{Hubert_thesis}. A feature of the $\beta^*$ = 90 m optics is that $L_y \gg L_x$ and hence $t_y$ is determined with almost an order of magnitude better precision than $t_x$.  For central diffraction, the diffractive mass can be reconstructed from
the $\xi$ measurements of the two protons according to \vspace*{-0.15cm}
\begin{equation}
M^{2} = \xi_{1} \, \xi_{2} \, s \:. \label{eqn_DPEmass} \vspace*{-0.25cm}
\end{equation}
The mass resolution for central diffractive events at different optics is also quoted in Table~\ref{tab_pres}. If the
scattering vertex is determined with high precision ($\sim$ 30\,$\mu$m) with the CMS tracking detectors during common data
taking, a substantial improvement in the $\xi$ and $M$ measurement is achieved at $\beta^*$ = 90 m.

\section{TOTEM physics programme}
Given its unique coverage for charged particles at high rapidities,
TOTEM is ideal for studies of forward phenomena, including elastic
and diffractive scattering.
Its main physics goals, precise measurements of the total cross-section
and of elastic scattering over a large range in $|t|$,
are of primary importance for distinguishing between
different models of soft $pp$ interactions. Furthermore, as energy flow and particle
multiplicity of inelastic events peak in the forward region, the large rapidity
coverage and proton detection on both sides allow the study of a wide range
of processes in inelastic and diffractive interactions.

\subsection{Elastic scattering and diffraction}
\label{sec:elastic}
\begin{figure}[h!]
\vspace*{-0.8cm}
\begin{center}
\mbox{
\epsfig{file=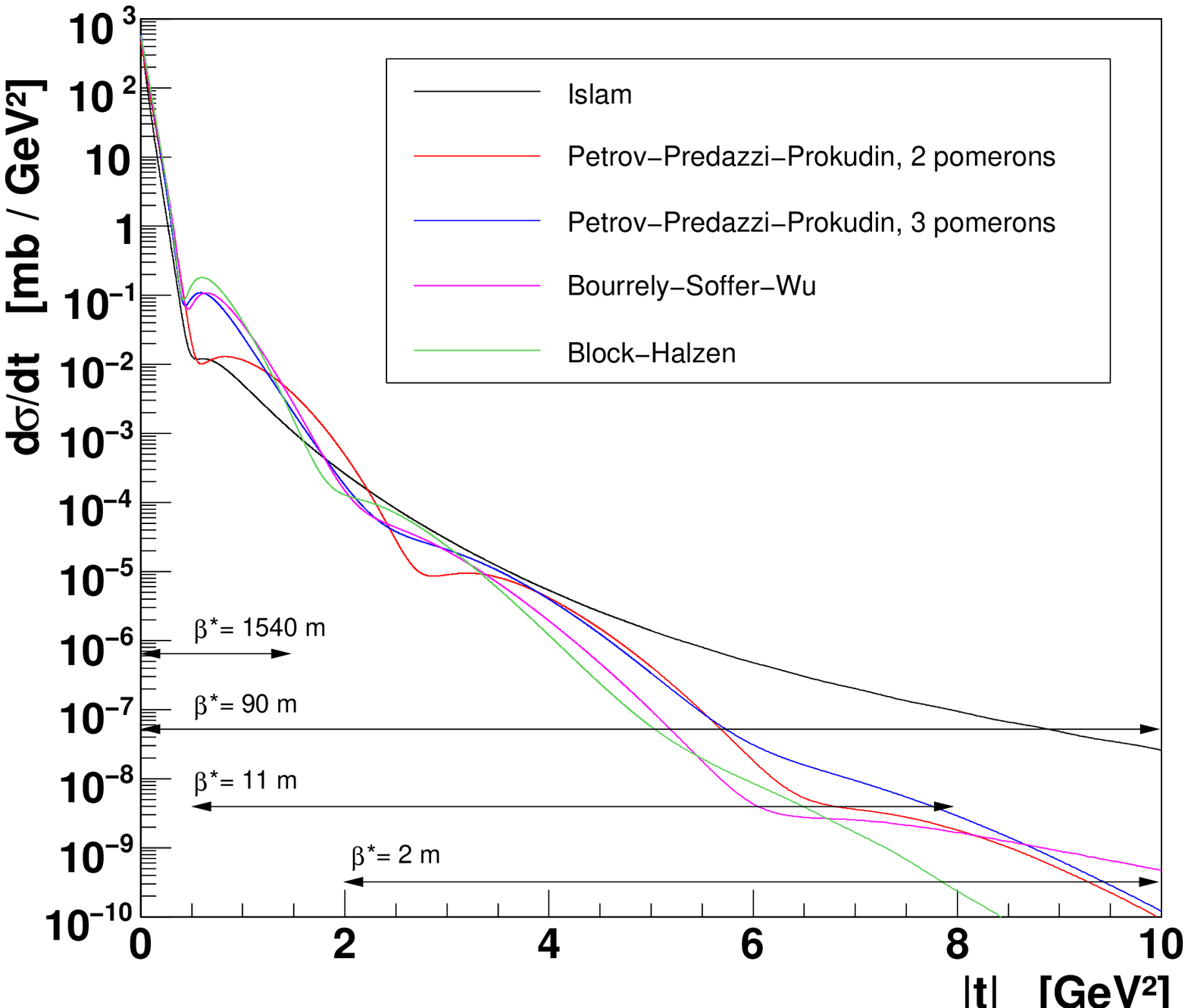,height=6.0cm}
\hspace*{2mm}
\epsfig{file=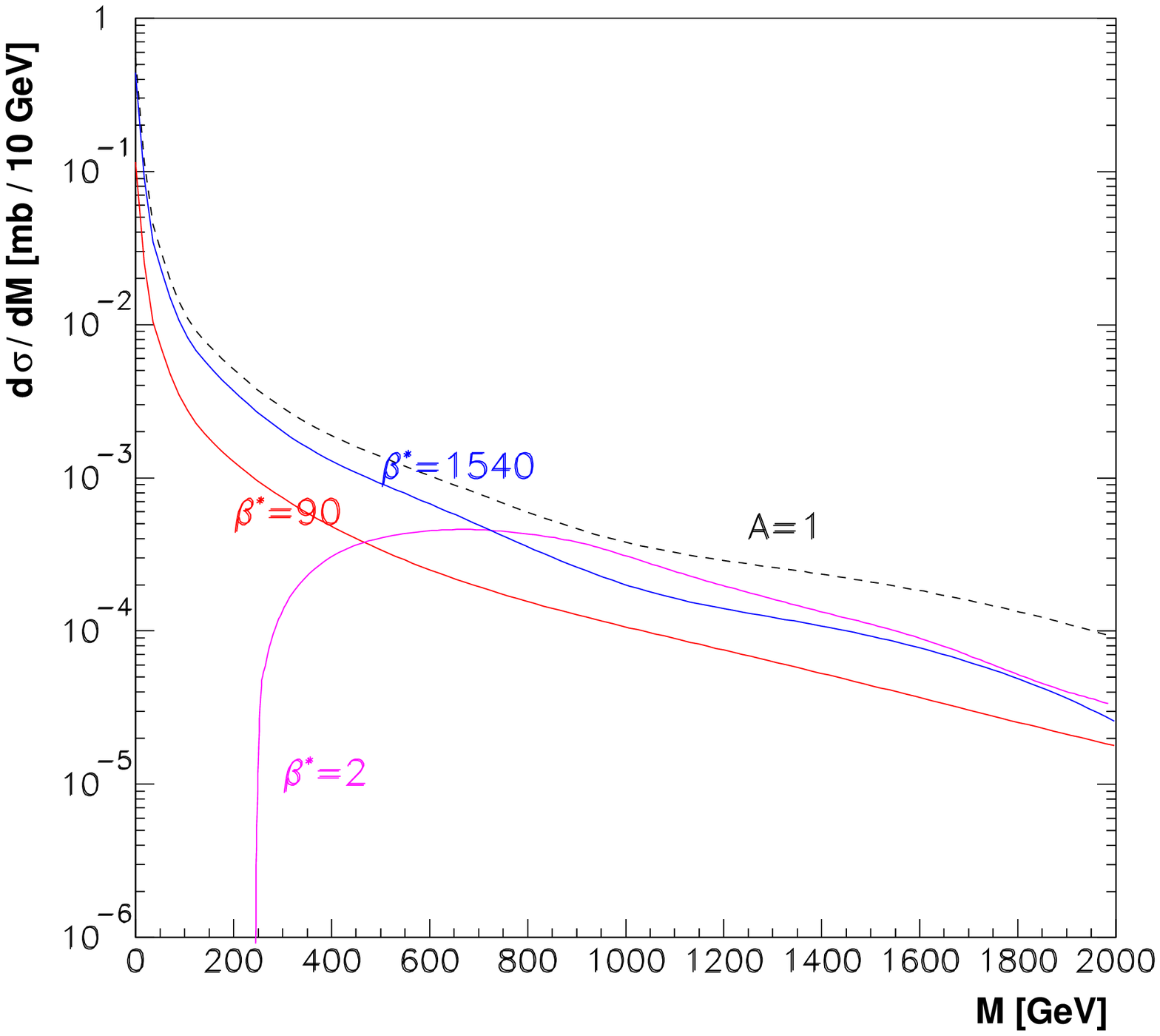,height=6.0cm}
}
\vspace*{-0.5cm}
\end{center}
\caption{
Left: differential cross-section of elastic scattering at $\sqrt{s}$ = 14 TeV
as predicted by various
models together with the $t$-acceptance ranges of different optics configurations.
Right: predicted differential cross-section of central diffraction at
$\sqrt{s} = 14$\,TeV with (solid) and
without (dashed) taking the proton acceptance into account for different
optics configurations.}
\label{fig_physperf_crosssect_accept}
\end{figure}

Much of the interest in large-impact-parameter collisions centers on elastic
scattering and soft inelastic diffraction. The differential
cross-section of elastic $pp$ interactions at 14\,TeV, as predicted by different
models~\cite{elasticmodel1, elasticmodel2, elasticmodel3, elasticmodel4}, is given in
Figure~\ref{fig_physperf_crosssect_accept} (left).
Increasing $|t|$ means looking deeper into the proton at smaller distances.
Several $|t|$-regions with different behavior (at $\sqrt{s}$ = 14\,TeV) can be distinguished:
\begin{itemize}
\item $|t| < 6.5\cdot 10^{-4}\,{\rm GeV}^{2}$: The Coulomb region dominated by photon
exchange: $d\sigma / dt \sim 1 / t^{2}$.
\item $10^{-3}\,{\rm GeV}^{2} < |t| < 0.5\,{\rm GeV}^{2}$:
The nuclear region, described in a simplified way
by "single-Pomeron exchange": $d\sigma / dt \sim {\rm e}^{-B\,t}$, is crucial for the extrapolation of
the differential counting-rate $dN_{el}/dt$ to $t=0$, needed for
the luminosity-independent total cross-section measurement.
\item 0.5\,GeV$^{2} < |t| < 1$\,GeV$^{2}$:
A region exhibiting the diffractive structure of the proton.
\item $|t| > 1$\,GeV$^{2}$:
Domain of central elastic collisions,
described by perturbative QCD, e.g. via
triple-gluon exchange with a predicted cross-section $\propto |t|^{-8}$. The model dependence of the predictions being very pronounced in this region, measurements will test the validity of different models.
\end{itemize}
TOTEM will cover the full elastic
$|t|$-range from 0.002 up to 10 GeV$^{2}$ by combining data from runs at several optics configurations as indicated in Figure~\ref{fig_physperf_crosssect_accept} (left). With typical expected LHC machine cycle times of $10^{4}-10^{5}$ s, enough statistics at low $|t|$ values
can be accumulated in one run. This statistics is also sufficient
for track-based alignment of the RP
detectors. The overlap between the acceptances
of the different optics configurations will allow for cross-checks of the
measurements.

\begin{figure}[h!]
\vspace*{-0.8cm}
\begin{center} \epsfig{file=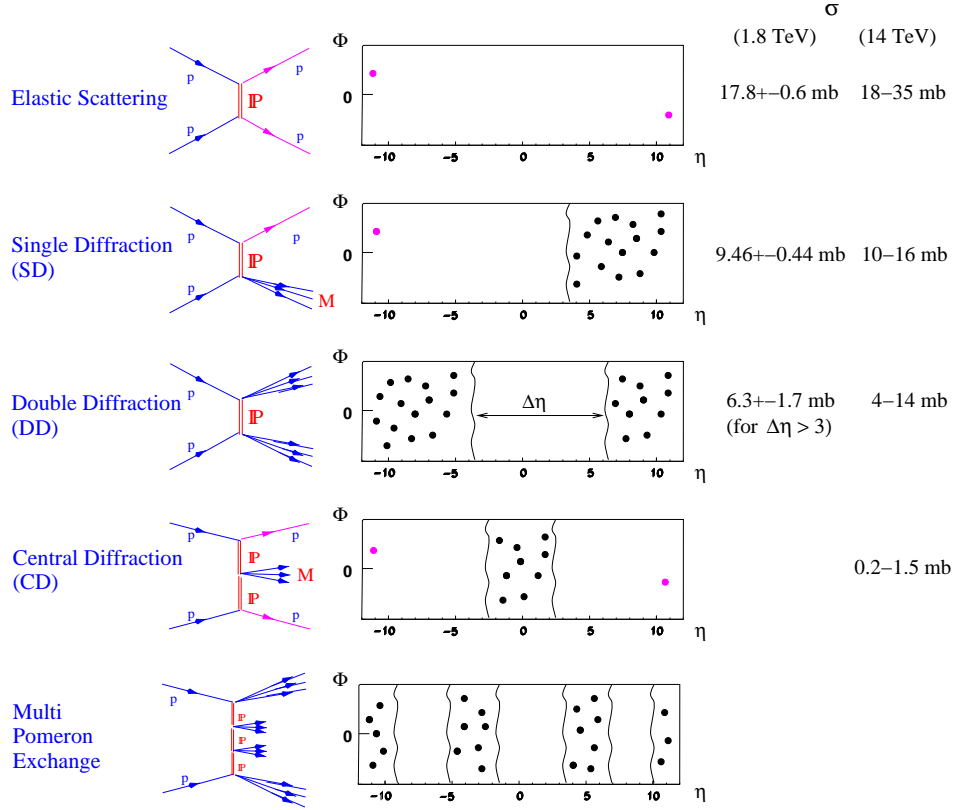,width=12.5cm}
\end{center}
\caption{Different classes of diffractive processes and their cross-sections as measured
at Tevatron and as estimated for the LHC.}
\label{fig_eventclasses}
\end{figure}

Diffractive scattering comprises single diffraction, double diffraction,
central diffraction (a.k.a. ``double Pomeron exchange''), and higher order
(``multi Pomeron'') processes, shown in Figure~\ref{fig_eventclasses} with their cross-sections as measured at Tevatron~\cite{cdf_elastic, E811_elastic, cdf_sd, cdf_dd} and
as predicted for LHC~\cite{elasticmodel1, elasticmodel2, elasticmodel3, elasticmodel4, khoze_sigmatot, maor_sigmatot, phojet}. Together with elastic scattering these
processes represent about 50\,\% of the total cross-section.
Many details of these processes with close ties to proton
structure and low-energy QCD are still poorly understood.
Majority of diffractive events (Figure~\ref{fig_eventclasses}) exhibits
intact (``leading'') protons in the final state, characterized by
their $t$ and $\xi$. For large $\beta^*$
(see Figure~\ref{fig_physperf_accept}, right) most of these protons
can be detected in the RP detectors.
Already at an early stage, TOTEM
will be able to measure $\xi$-, $t$- and mass-distributions in soft
central and single diffractive events.
The full structure of diffractive events with one or more sizeable rapidity
gap in the particle distribution (Figure~\ref{fig_eventclasses}) will be
optimally accessible
when the detectors of CMS and TOTEM will be
combined for common data taking with an unprecedented rapidity coverage, as
discussed in~\cite{physicstdr}.

Figure~\ref{fig_physperf_crosssect_accept} (right) shows the
predicted central diffractive mass distribution~\cite{phojet} together with the acceptance
corrected distributions for three
different optics. With
high and intermediate $\beta^{*}$ optics, all diffractive mass values are observable.
For low $\beta^{*}$ optics on the other hand, the acceptance starts at $\sim$ 250\,GeV but higher statistics for high masses will be collected due to the
larger luminosity.
By combining data from runs at low $\beta^{*}$ with data from high or
intermediate $\beta^*$ runs,
the differential cross-section as function of the central diffractive mass
can be measured with good precision
over the full mass range.

\subsection{Total $pp$ cross-section}
 The optical theorem relates the total $pp$ cross-section $\sigma_{tot}$ and the luminosity $\mathcal{L}$ to the differential elastic counting-rate $dN_{el}/dt$ at $t = 0$ and the total elastic $N_{el}$ and inelastic $N_{inel}$ rates as:  \vspace*{-0.10cm}
\begin{eqnarray}
\sigma_{tot} = \frac{16 \pi}{1 + \rho^{2}} \cdot
\frac{dN_{el}/dt |_{t=0}}{N_{el} + N_{inel}}\: \, \, \, \, \, & {\rm and} &
\, \, \, \, \,  \mathcal{L} = \frac{1 + \rho^{2}}{16 \pi} \cdot
\frac{(N_{el} + N_{inel})^{2}}{dN_{el}/dt |_{t=0}}\: .\label{eqn_tot}   \vspace*{-0.25cm}
\end{eqnarray}
The parameter $\rho = \mathcal{R}[f_{el}(0)]/\mathcal{I}[f_{el}(0)]$,
where $f_{el}(0)$ is the forward nuclear elastic amplitude,
has to be taken from theoretical
predictions. Since $\rho \sim$ 0.14 enters only as a $1+ \rho^2$ term, its impact is small. The extrapolation of existing $\sigma_{tot}$ measurements to LHC energies leaves a wide range for the expected value of $\sigma_{tot}$ at LHC, typically between 85 and 120 mb, depending on the model used for the extrapolation. TOTEM aims at a 1 \% $\sigma_{tot}$ measurement.
Hence the quantities to be measured are the following:
\begin{itemize}
\item the inelastic rate $N_{inel}$ consisting of
both non-diffractive minimum bias events and
diffractive events,
which can almost completely be measured by T1, T2 and the RP detectors;
\item the total nuclear elastic rate $N_{el}$ measured exclusively by the RP
system;
\item $dN_{el}/dt |_{t=0}$: the nuclear part of the elastic cross-section
extrapolated to $t = 0$. \end{itemize} \noindent A summary of the uncertainties on $\sigma_{tot}$ at different high $\beta^*$ optics configurations is given in Table~\ref{tab_error_sigmatot}. Here only the main uncertainties are described. The extrapolation procedure and uncertainty estimates are described in more detailed in Ref.~\cite{LHCarticle}. At $\beta^*$ = 90 m, protons with
$|t| > 0.03$ GeV$^{2}$ are observed, whereas $|t|_{min}$ = 10$^{-3}$ GeV$^{2}$ at
$\beta^*$ = 1540 m, leading to a significantly smaller uncertainty contribution due to $N_{el}$, 0.1 \% compared to 2 \%, and to the extrapolation of $dN_{el}/dt$ to $t = 0$, 0.2\% compared to 4\%.

\begin{table}[ht]
\begin{center}
\begin{tabular}{|ccc|}\hline
Uncertainty & $\beta^*$ = 90 m & $\beta^*$ = 1540 m \\ \hline
Extrapolation of $dN_{el}/dt$ to $t = 0$ & $\pm$ 4 \% & $\pm$ 0.2 \% \\
Elastic rate $N_{el}$ & $\pm$ 2 \% & $\pm$ 0.1 \% \\
Inelastic rate $N_{inel}$ & $\pm$ 1 \% & $\pm$ 0.8 \% \\
$\rho$ parameter & $\pm$ 1.2 \% & $\pm$ 1.2 \% \\ \hline
Total $\sigma_{tot}$ & $\pm$ 4--5 \% &  $\pm$ 1--2 \% \\
\hline
\end{tabular}
\caption{Relative uncertainty on the total $pp$ cross-section $\sigma_{tot}$ measurement estimated at different high $\beta^*$ optics configurations. Note that the total uncertainty takes into account the correlations between the uncertainties, notably the strong correlation between the extrapolation of the differential elastic counting-rate $dN_{el}/dt$ to $t = 0$ and the elastic rate $N_{el}$.}
\label{tab_error_sigmatot}
\end{center} \vspace*{-0.8cm}
\end{table}

The largest contribution to the uncertainty on $\sigma_{tot}$ at
$\beta^*$ = 90 m comes from the extrapolation of $dN_{el}/dt$ to $t = 0$; mainly due to systematics in the $t$-measurement from uncertainties in $L$ and $v$ (see Eq.\ref{eqn_coord}). This contribution will be reduced to 0.1 \% at $\beta^*$ = 1540 m requiring, however, an improved knowledge of $L$ and $v$ and a RP alignment precision of better than 50 $\mu$m. The
dominating uncertainty, 0.2 \%, will then be due to the model-dependent extrapolation procedure.
For $\beta^*$ = 1540 m, the largest contribution to the $\sigma_{tot}$ uncertainty will most likely come from $N_{inel}$, mainly from trigger losses in single and double diffractive
events. The lost events, corresponding to $\sim$ 3\,mb, have very low diffractive mass $M$
(below $\sim$ 10\,GeV/c$^2$). As a consequence, all particles have pseudo-rapidities beyond the T2 acceptance and hence escape detection of the single arm trigger.
To obtain the total inelastic rate, the fraction of events lost due to
the incomplete angular coverage is estimated by extrapolating the reconstructed $1/M^2$ distribution. The uncertainty on $N_{inel}$ after corrections is estimated to be 0.8 and 1 \% for $\beta^*$ = 1540 and 90 m optics, respectively. The uncertainty on the $\rho$ parameter as estimated from lower energy measurement~\cite{compete} gives a $\sigma_{tot}$ uncertainty of 1.3 \%. A reduction is expected when $\rho$ is measured at the LHC via the
interference between Coulomb and nuclear contributions to the elastic
scattering cross-section ~\cite{blois2005}.

At an early stage in 2009 with non-optimal beams and $\beta^*$ = 90 m, TOTEM will measure $\sigma_{tot}$ ($\mathcal{L}$) with a 4--5 \% (7 \%)
relative precision. After having
understood the initial measurements and with improved beams at
$\beta^{*} = 1540\,$m, a final
relative
precision on $\sigma_{tot}$ ($\mathcal{L}$) of 1 \% (2\%) should be achievable.

\section{Early physics with TOTEM}

The early runs at the LHC start will be characterized by low $\beta^*$ beams with a reduced number of
bunches and a lower number of protons per bunch. Under these conditions diffractive protons in the $\xi$ range of 0.02 - 0.2 will be within the
 acceptance of RP220 giving TOTEM ample opportunities to make first soft
diffractive studies. The early physics programme of TOTEM in stand-alone runs will concentrate on measurements of individual cross-sections and event topologies for the following processes:
\begin{itemize}
 \item central diffractive events with diffractive masses between $\sim$ 250 GeV and $\sim$ 2.8 TeV;
\item single diffractive events with diffractive masses between $\sim$ 2 TeV and $\sim$ 6 TeV;
 \item elastic scattering events with $|t|$ values between $\sim$ 2 GeV$^2$ and $\sim$ 10 GeV$^2$;
 \item forward charged particle multiplicity of inelastic $pp$ events in the 3.1 $\le |\eta| \le$ 6.5 region.
\end{itemize}

\noindent
The cross-sections for the above processes are large ($\gtrsim$ 5 $\mu$b) even if the TOTEM acceptance is included, with the exception of high-$|t|$ elastic scattering. As an example, the BSW model~\cite{elasticmodel2} predicts an integrated elastic cross-section of $\sim$ 60 nb for
$|t| >$ 2 GeV$^2$.  This prediction, together with the predictions of Ref.~\cite{khoze_sigmatot, maor_sigmatot, phojet}, imply that for an
integrated luminosity of $\sim$ 10 pb$^{-1}$, TOTEM would collect more than 10$^7$ central and 10$^8$ single diffractive events, together with $\sim$ 10$^5$ high-$|t|$ elastic events allowing a first test of the validity of different models as discussed in section~\ref{sec:elastic}. The main background to diffractive events at low $\beta^*$ is either due to two overlapping $pp$ collisions, like e.g. two overlapping single diffractive events for central diffraction, or one $pp$ collision overlapping with beam induced proton background. Hence the event purity will depend strongly on the average number of $pp$ collisions per bunch crossing, which should be significantly smaller than one. The beam induced proton background not due to $pp$ collisions in IP5 will be studied in bunch crossings where normal bunches meet "empty" bunches. The interest in the forward charged particle multiplicity is two-fold: first as a basic measurement of $pp$ interaction at LHC energies and secondly as valuable input to the modeling of very high energy cosmic rays\cite{physicstdr}.

The installation schedule of the TOTEM detectors depends crucially on the CMS  installation schedule as well as on the LHC commissioning schedule. The full experiment is planned to be installed for the 2009 LHC running. The focus in the early LHC runs will be to understand the performance of the detectors and other vital parts like trigger and data acquisition, especially the approach of the RP detectors to the beam. The feasibility and time scale of the early physics programme will critically depend on the LHC performance in terms of luminosity and beam induced background in the TOTEM detectors.

\section{Diffractive proton detection at IP3}
It has been suggested that the central exclusive diffractive process \vspace*{-0.20cm}
\begin{equation}
p \, p \, {\rightarrow} \, p \, + \, X \, + \, p \, \, ,
\label{excluproc}   \vspace*{-0.25cm}
\end{equation}
where a "+" denotes a rapidity gap, could complement the standard methods of searching and studying new particles ("X") at LHC, see e.g. Ref.~\cite{khoze}. The main advantage is that
the mass of the centrally produced particle $X$ can be reconstructed from the measured $\xi$ values of the outgoing protons as shown in Eq.~\ref{eqn_DPEmass}. Provided that the two $\xi$ values can be determined with sufficient precision, peaks corresponding to particle resonances may appear in the reconstructed diffractive mass distribution independent of the particles' decay modes. These measurements should be performed with high luminosity optics since the cross-sections are expected to be small. The work presented here aims to find the best detector locations at LHC in terms of $\xi$ acceptance and resolution for the proton measurement in central diffractive events.

The diffractive proton acceptance of near beam detectors
is determined by the ratio
$D_{x}/\sigma_{x}$ between horizontal dispersion and beam width. With larger $D_{x}$
the protons are deflected further away from the beam center, while
the closest safe approach of a detector to the beam is given by a multiple -- typically 10 to 15 -- of $\sigma_{x}$.
By construction, the LHC region where $D_{x}$ and
$D_{x}/\sigma_{x}$ are maximized and hence the sensitivity to particle $X$, is the momentum cleaning insertion in IP3,
where off-momentum beam protons are intercepted.
The idea is to install proton detectors pairs with a lever arm of several tens of meters close to IP3 to detect diffractive protons in both beams just before they are absorbed by the momentum cleaning collimators. In addition to promising perspectives in diffraction, the placement of detectors
in front of the collimators has advantages for accelerator diagnostics and protection. The technical aspects of placing proton detectors at IP3 is being worked out together with the LHC collimation group.

\begin{figure}[h!]
\begin{center}
\vspace*{-0.3cm}
\mbox{\hspace*{-0.5cm}
\epsfig{file=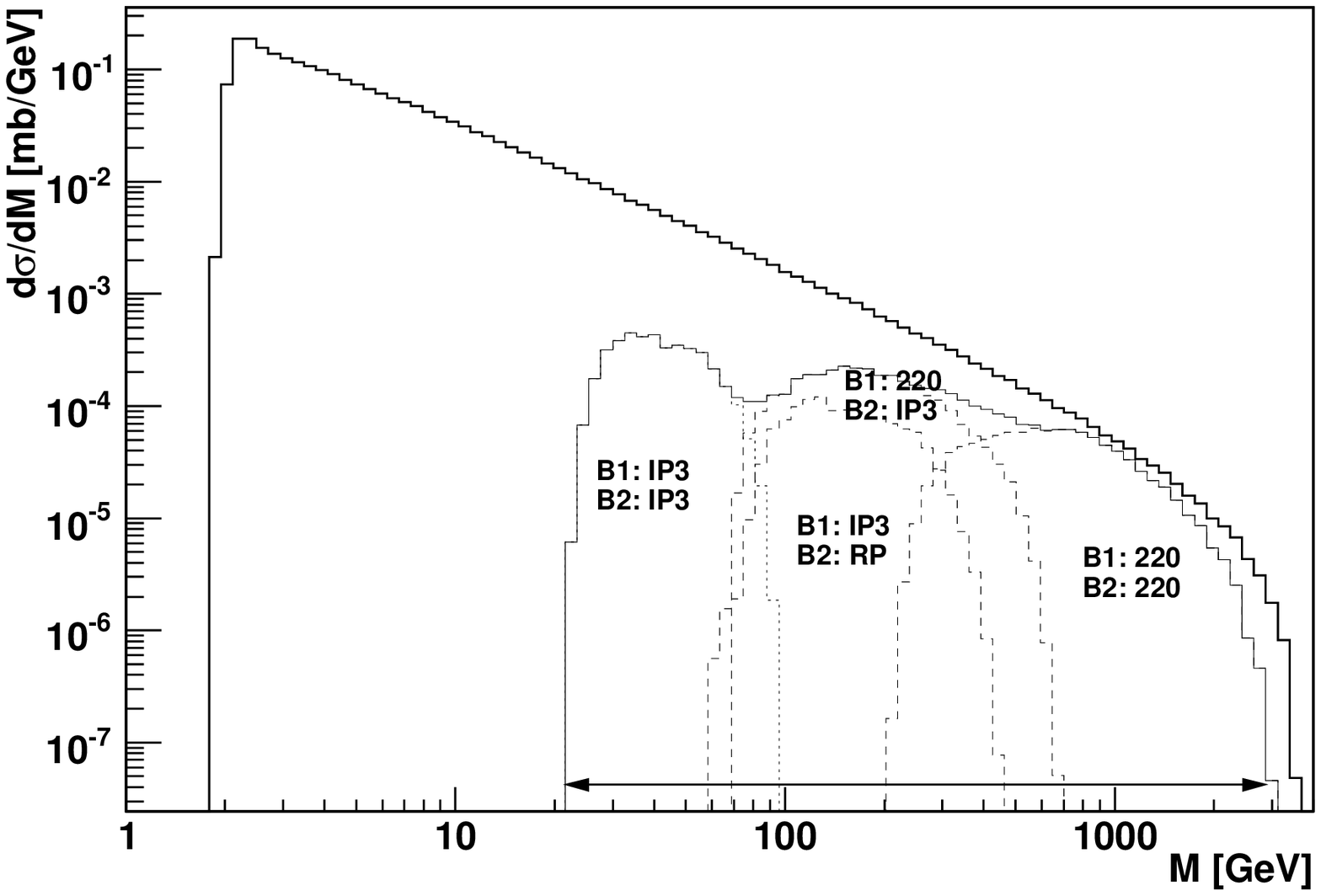,height=6.1cm}
\epsfig{file=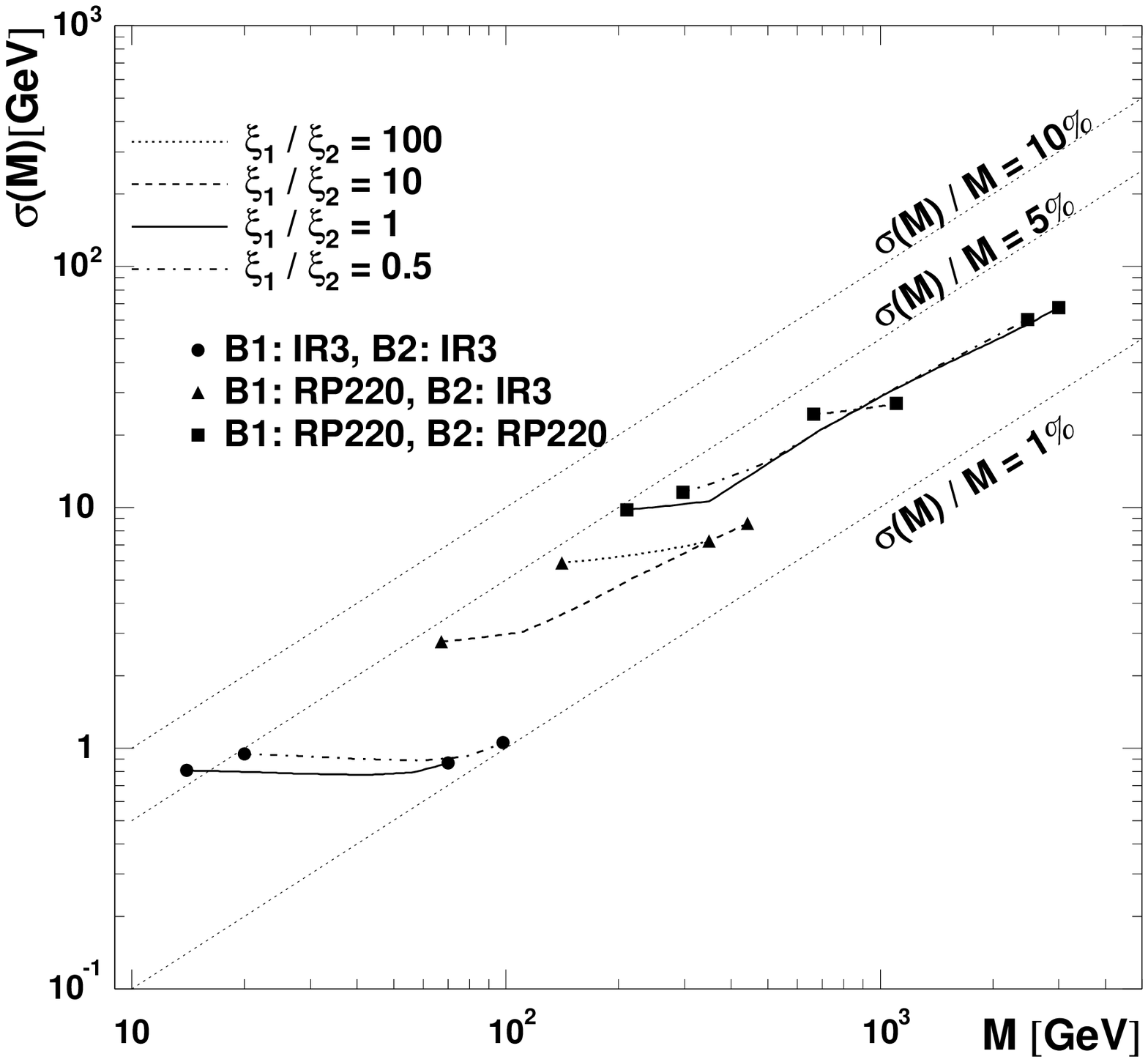,height=6.1cm}}
\vspace*{-0.5cm}
\end{center}
\caption{Left: predicted diffractive mass distribution for central diffractive events with events indicated that have both protons within the acceptance of different combinations of the IP3 and RP220 detectors. Right: mass resolution for central diffractive events for some $\xi_1 / \xi_2$ ratios. B1 and B2 refers to protons turning clockwise and anticlockwise, respectively, in the LHC.}
\label{fig_ip3}
\end{figure}

The proton acceptance and resolution of an experiment with detectors at the
TOTEM RP220 location and at IP3
have been studied~\cite{Karsten_talk} by fully tracking the protons along the LHC ring
with the MAD-X~\cite{madx} program using standard LHC high luminosity $\beta^*$ = 0.55 optics. The detector acceptance at IP3 for protons originating from diffractive scattering in IP5 is
$0.0016 \le \xi \le 0.004$ and $0.0016 \le \xi \le 0.01$
for protons turning clockwise ("B1") and anticlockwise ("B2") in the LHC, respectively. This complements well the $0.02 \le \xi \le 0.20$ acceptance of RP220 for both beams. \mbox{The IP3 acceptance for B1 protons
is} reduced since these protons have to pass through the aperture limiting betatron cleaning insertion at IP7. In case of central diffraction~\cite{phojet}, this gives access to diffractive masses from 25\,GeV to 2.8\,TeV as shown in Fig.~\ref{fig_ip3} (left).
A $\xi$ resolution $\leq$ 10$^{-4}$ for protons detected at IP3 is obtained in the study implying that the resolution will be limited by the beam energy spread of 1.1 $\cdot$ 10$^{-4}$. Combined with protons detected at RP220, this leads to a relative mass resolution ranging between 1 and 5 \% for central diffractive events over the whole mass range as shown in Fig.~\ref{fig_ip3} (right). The mass resolution depends on the ratio $\xi_1 / \xi_2$, where $\xi_1$ and $\xi_2$ are the $\xi$ value of the clockwise and anticlockwise turning proton, respectively.

The protons detectors at IP3 would in fact see diffractive protons with similar acceptance from all LHC interaction points (IP) and could by measuring the difference of the proton arrival times determine at which IP the event occurred. This way the low mass central diffractive spectrum could be determined independently for each IP and be used as means of an inter-experimental luminosity calibration.

\section{Summary}
The TOTEM physics program aims at a deeper understanding of the proton structure by measuring the total and elastic $pp$ cross
sections and by studying a comprehensive menu of diffractive processes. TOTEM will run under all LHC beam conditions to maximize the coverage of the studied processes.
Special high ${\beta}^*$ runs are needed for the total $pp$ cross section measurement with the luminosity-independent method and for soft diffraction with large forward proton acceptances. At an early stage with non-optimal beams and an intermediate $\beta^*$, TOTEM will measure $\sigma_{tot}$ with a 4--5 \% precision. With improved understanding of the beams and
$\beta^{*} = 1540\,$m, a precision on $\sigma_{tot}$ of 1\% should be achievable.
The measurement of elastic scattering in the range
$10^{-3} < |t| < 10$ GeV$^{2}$ will allow to distinguish among a wide
range of predictions according to current theoretical models.
Early low $\beta^*$ runs will provide first opportunities for measurements of soft diffraction for masses above $\sim$250 GeV and $\sim$2 TeV in central and single diffractive events, respectively, as well as studies of the forward charged multiplicity in inelastic $pp$ events. Having proton detectors at IP3 would highly extend the diffractive mass acceptance of TOTEM for high luminosity runs giving e.g. a continuous mass acceptance from 25 GeV to 2.8 TeV for central diffractive events.
Finally, hard diffraction as well as many forward physics subject will be studied in collaboration with CMS taking advantage of the unprecedented rapidity coverage for charged and neutral particles.

\bibliographystyle{heralhc}
{\raggedright
\bibliography{TOTEM_heralhc}
}
\end{document}